\documentclass[usenatbib,fleqn,12pt]{wlscirep}

\usepackage{graphicx}
\usepackage{amssymb}
\usepackage{newfloat}
\DeclareFloatingEnvironment[name={Extended Data Figure}]{extfig}

\newcommand{\apj}{Astrophys. J.}           
\newcommand{\apjl}{Astrophys. J.}           
\newcommand{\mnras}{Mon. Not. R. Astron. Soc.}       
\newcommand{\nat}{Nature}
\newcommand{\aap}{Astron. Astrophys.}
\newcommand{\araa}{Annual Rev. Astron. Astrophys.}
\newcommand{\aj}{Astron. J.}
\newcommand{\pasp}{Pubbl. Astron. Soc. Pacific}

\title{The recurrent impact of the Sagittarius dwarf galaxy on the star formation history of the Milky Way disc}

\author[1,2,*]{Tom\'as Ruiz-Lara}
\author[1,2]{Carme Gallart}
\author[3]{Edouard J. Bernard}
\author[4,5]{Santi Cassisi}

\affil[1]{Instituto de Astrof\'isica de Canarias, E-38200 La Laguna, Tenerife, Spain}
\affil[2]{Departamento de Astrof\'isica, Universidad de La Laguna, E-38205 La Laguna, Tenerife, Spain}
\affil[3]{Universit\'e C\^ote d'Azur, OCA, CNRS, Lagrange, France}
\affil[4]{INAF -- Astronomical Observatory of Abruzzo, via M. Maggini, sn, 64100 Teramo, Italy}
\affil[5]{INFN, Sezione di Pisa, Largo Pontecorvo 3, 56127 Pisa, Italy}
\affil[*]{tomasruizlara@gmail.com}

\begin{abstract}

\end{abstract}

\begin{document}

\flushbottom
\maketitle

\thispagestyle{empty}
\vspace{-1.7cm}

{\bf
\noindent
Satellites orbiting disc galaxies can induce phase space features such as spirality, vertical heating and phase-mixing in their discs\cite{quinn1993, velazquez1999, quillen2009, purcell2011}. Such features have also been observed in our own Galaxy\cite{eggen1969, siebert2011, williams2013}, but the complexity of the Milky Way disc has only recently been fully mapped thanks to Gaia DR2 data\cite{brown2018, katz2018}. This complex behaviour is ascribed to repeated perturbations induced by the Sagittarius dwarf galaxy (Sgr\cite{ibata1994}) along its orbit\cite{delavega2015, antoja2018, laporte2018, laporte2019}, pointing to this satellite as the main dynamical architect of the Milky Way disc. Here, we model Gaia DR2 observed colour-magnitude diagrams to obtain the first detailed star formation history of the $\sim$ 2-kpc bubble around the Sun. It reveals three conspicuous and narrow episodes of enhanced star formation that we can precisely date as having occurred 5.7, 1.9 and 1 Gyr ago. Interestingly, the timing of these episodes coincides with proposed Sgr pericentre passages according to i) orbit simulations\cite{law2010, laporte2018}, ii) phase space features in the Galactic disc\cite{gomez2013, delavega2015, antoja2018, laporte2019}, and iii) Sgr stellar content\cite{siegel2007, deboer2015}. These findings most likely suggest that Sgr has also been an important actor in the build-up of the Milky Way disc stellar mass, with its perturbations repeatedly triggering major episodes of star formation. The high mass ratio involved, the global rather than centrally concentrated nature of the star formation enhancements\cite{mihos1994, hernquist1995, moreno2015}, and the precisely known characteristics of the interaction parameters of Sgr with the Milky Way, set important constraints on hydrodynamical simulations\cite{teyssier2010, chien2010, moster2011} of interaction-induced star formation in galaxies.
}

\vspace{0.5cm}

\begin{figure}[h]
\begin{center}
\includegraphics[width=0.9\textwidth]{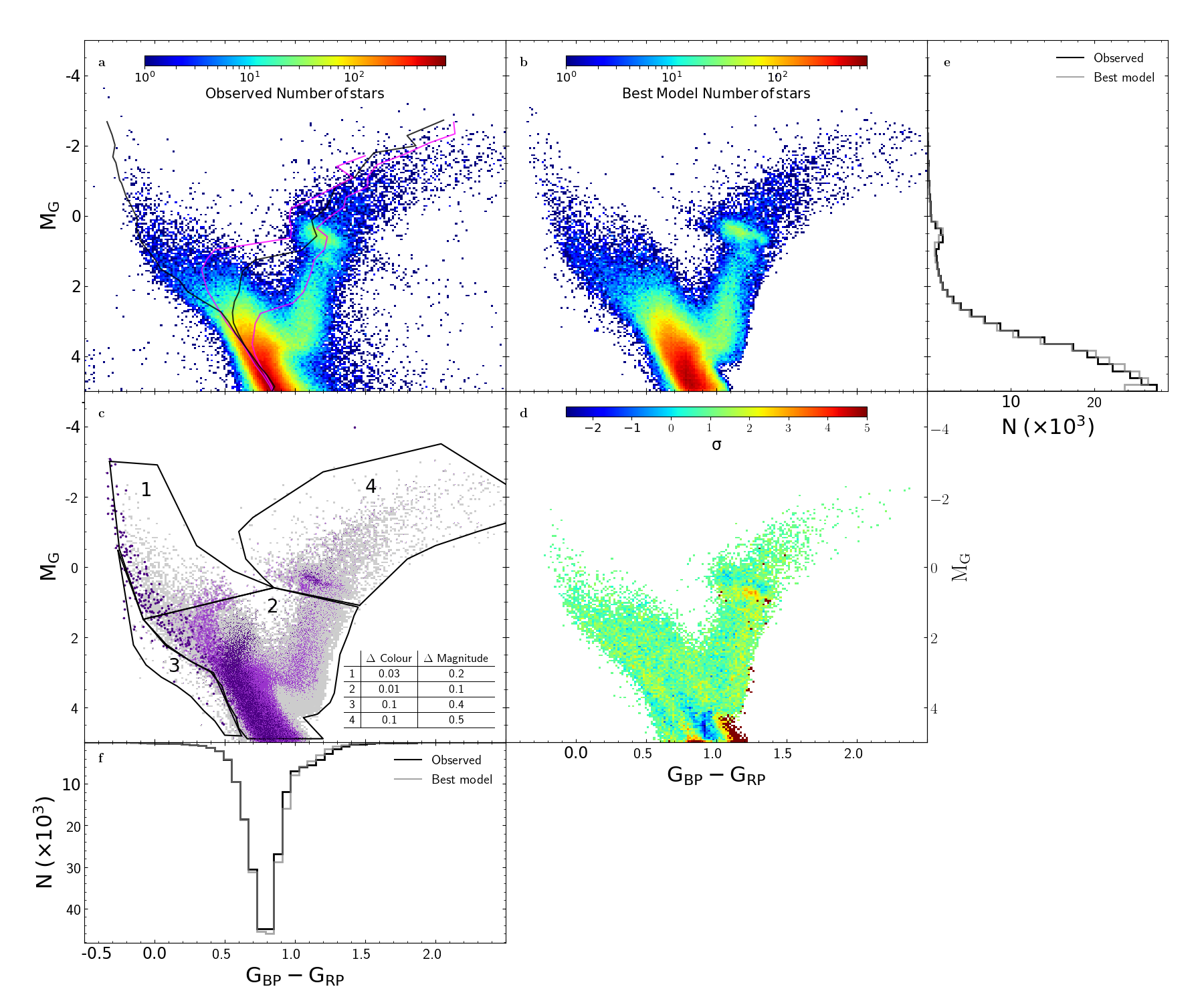} 
\end{center}
\caption{{\bf Colour-magnitude diagram of the $\sim$ 2-kpc bubble around the Sun.} {\bf a} Observed M$_{\rm G}$~vs.~G$_{\rm BP}$-G$_{\rm RP}$ Gaia DR2 CMD using derredened magnitudes. Lines are representative of the average run in the colour-magnitude plane of theoretical stars belonging to each star formation enhancement. {\bf b} Best fitting model to the observed CMD. In both cases we use a Hess diagram representation of the CMDs with the number of stars in logarithmic scale. {\bf c} Position in the CMD of synthetic stars linked to each of the four (including also the youngest one) star formation enhancements using different hues of purple. The {\it bundle} strategy chosen for the CMD fitting is overplot in black. The sizes of the boxes in each bundle is given as an inset table. Note that the most weight in the fitting corresponds to the more densely sampled bundle 2. {\bf d} Residual difference between the observed and the best model CMDs in poissonian sigmas. Note that in the great majority of the CMD the residuals are lower than 2 sigmas. {\bf e-f} M$_{\rm G}$ and G$_{\rm BP}$-G$_{\rm RP}$ distributions of the observed (black) and best model (gray) stars, respectively.}
\label{fig:plot1}
\end{figure}

\begin{figure}[h]
\begin{center}
\includegraphics[width=0.7\textwidth]{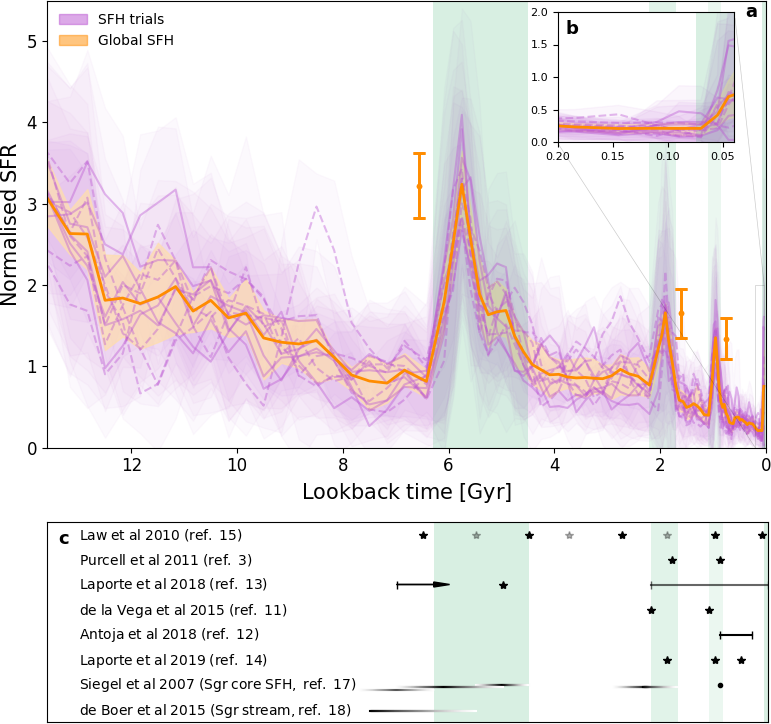} 
\end{center}
\caption{{\bf Star formation history representative of the $\sim$ 2 kpc bubble around the Sun.} {\bf a}, Normalised star formation rate as a function of lookback time for the different tests under analysis. {\bf b}, The inset shows a zoom-in to the youngest  star formation enhancement spanning the last $\sim$~100 Myr. Solid or dashed purple lines display the results from the different tests using two different extinction coefficients. The solid orange line depicts the overall star formation history averaging all the tests (see Methods). Shaded areas and orange error bars account for the errors on the SFH determinations and peak significance, respectively. These error bars do not account for possible degeneracies in the solution or systematic errors induced by the choice of input parameters of our method (see Methods). Green areas highlight the recovered star formation enhancements. {\bf c}, Comparison between our results and the knowledge regarding the Sgr orbit and stellar content from literature. Asterisks and line intervals correspond to proposed pericentric passages (grey asterisks for secondary pericentres). Lines in grey-scale mimic variable star formation from black (star forming) to white (no star formation) for the Sgr main body\cite{siegel2007} or stream\cite{deboer2015}.}
\label{fig:plot2}
\end{figure}

\begin{figure}[h]
\begin{center}
\includegraphics[width=0.95\textwidth]{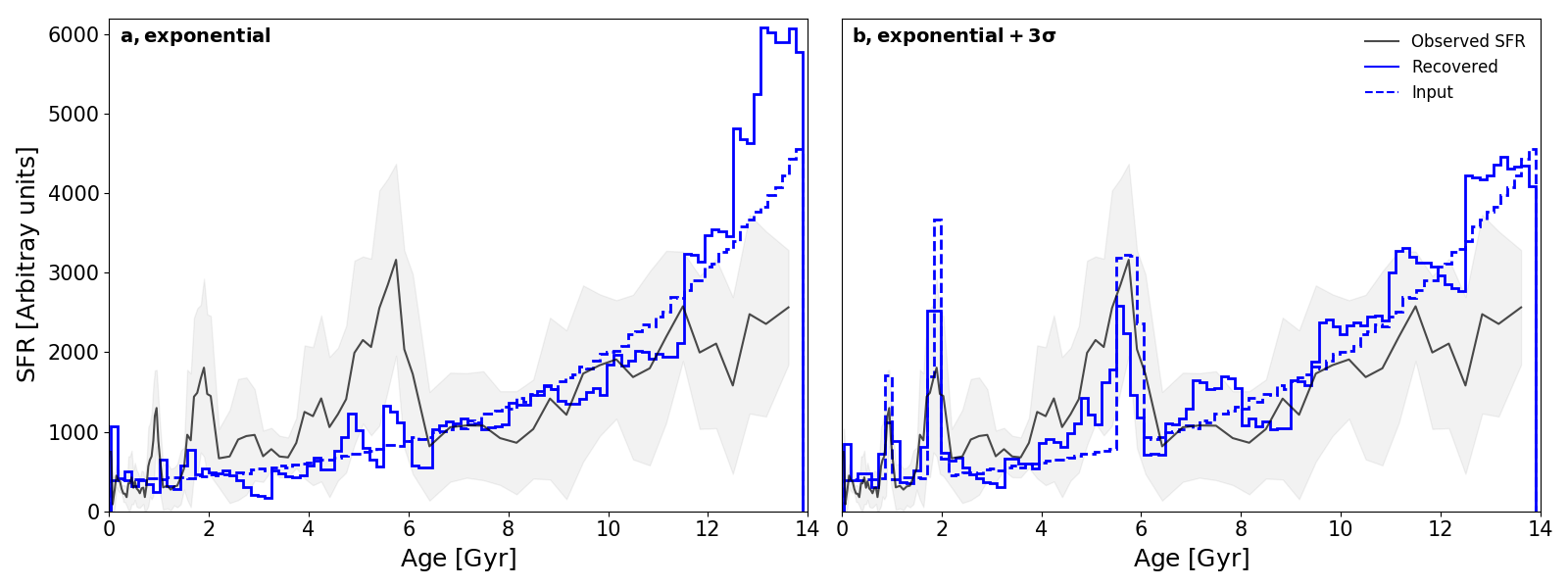} 
\end{center}
\caption{{\bf Testing the robustness of the SFH recovery}. In this figure we show the recovered SFH for the 2 kpc volume analysed from Gaia (grey),the  input SFH for the mock CMD (dashed blue line) and the recovered SFH after mimicking errors and applying our methodology (solid blue line). a: exponential decline. b: exponential decline plus a 3$\sigma$ burst.}
\label{fig:plot3}
\end{figure}

The accurate photometry and parallaxes delivered in Gaia DR2\cite{brown2018}, when combined with well-established techniques of modelling observed colour-magnitude diagrams\cite{gallart2005} (CMD, see Figure~\ref{fig:plot1}), enable the computation of unprecendently\cite{mor2019} detailed star formation histories (SFH) for large volumes within the Milky Way. In this work we use such techniques to characterise the stellar content within a bubble of $\sim$ 2-kpc radius around the Sun. Figure~\ref{fig:plot2} shows the derived SFH. We find that the Milky Way has been forming stars continuously along its entire history, with a decreasing intensity compatible with other massive spirals\cite{heavens2004}. On top of this overall behaviour, it displays three strikingly well-defined star formation enhancements that ocurred $\sim$~5.7, 1.9, and 1.0 Gyr ago, with a measured duration of 0.8, 0.2 and 0.1 Gyr, respectively (twice the $\sigma$ of a gaussian distribution fitted to each peak in the solution), and decreasing strength. There is also a hint of a fourth possible star formation burst spanning the last 70 Myr. Extensive testing using mock stellar populations \cite{hidalgo2011} on the ability of the method to recover star formation bursts in combination with an exponential decline confirms the detection of the bursts with a significance of 3-4$\sigma$ and above (see Figure~\ref{fig:plot3} and the method section). The episodic nature of the recovered SFH for the Milky Way disc is remarkable and, although hints of star forming bursts in the Milky Way disc at specific times have been suggested previously (see below), they have never been detected with such precision and significance as in the present work.

Previous determinations of the SFH of the Milky Way disk, using Hipparcos\cite{perryman1997} data\cite{hernandez2000, bertellinasi2001, vergely2002, cignoni2006} or different methodologies\cite{snaith2015, haywood2018b, mor2019, isern2019}, exist. However, they either concentrate on the solar neighbourhood within 250 pc and therefore provide a local view of the SFH of our Galaxy disc, or lack the age resolution needed to isolate individual star formation events. The combination of the large volume probed by the Gaia DR2 data and our state-of-the-art methodology allowed us to address both limitations. Such previous works show a relatively constant star formation rate with a clear enhancement in the last $\sim$ 4 Gyr \cite{vergely2002, cignoni2006, bernard2018proc, mor2019}. The differences between our derived SFH and that of previous works are considerably minimised when restricting our analysis to a similar region in the solar neighbourhood or when degrading our age resolution. However, while previous works are unable to detect an enhanced star formation around 5-6 Gyr ago, this is a robust feature of our SFH. We conclude that it is the combination of improved data (Gaia DR2) and an improved methodology that allows us to clearly detect this elusive (but important) star forming burst, previously barely hinted\cite{haywood2018b}. Our derived age-metallicity relation is also compatible with that found studying individual stars within the Milky Way\cite{feltzing2001, casagrande2011, bergemann2014}. 

\begin{figure}[h]
\begin{center}
\includegraphics[width=0.85\textwidth]{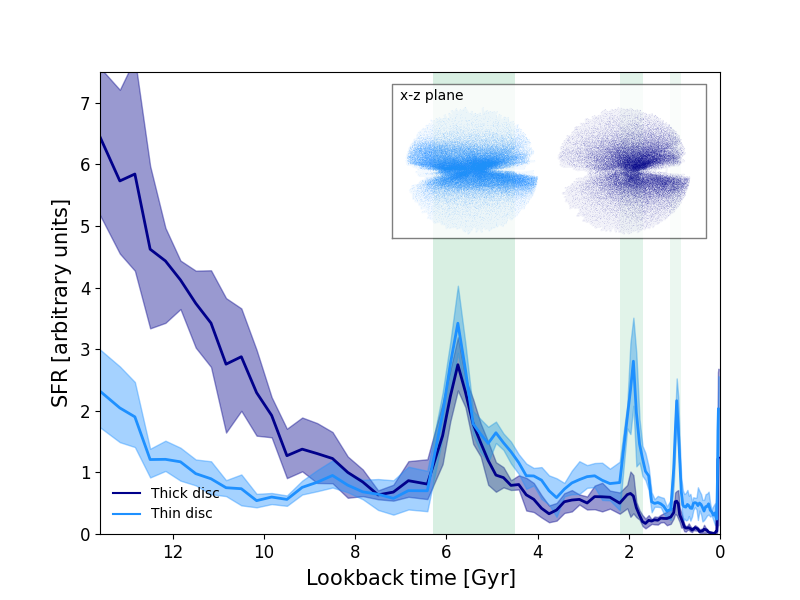} \\
\end{center}
\caption{Star formation history in the $\sim$ 2 kpc bubble around the Sun distinguishing between the (kinematically defined) thin and thick discs. Star formation rate (in arbitrary units) as a function of lookback time for the thin (light blue) and the thick (dark blue) discs. These SFHs have been obtained in a similar fashion to the orange line in figure~\ref{fig:plot2}. The inset shows the spatial distribution of the stars belonging to each substructure in the x-z plane. Some regions close to the Galactic plane have not been included in the analysis due to the high extinction in those directions. Stars have been divided into thin or thick disc stars following the kinematics criterion given in ref.~\cite{babusiaux2018} based on the tangential velocity.}
\label{fig:new_figure}
\end{figure}

The analysed volume (a bubble of $\sim$ 2 kpc of radius) is necessarily composed by stars of very different chemical and dynamical properties~\cite{hayden2015, katz2018}. In order to understand how the SFH varies for different kinematic components of the Milky Way disk, we assigned stars in the original sample to the kinematic thin and thick discs (thin and thick, hereafter, for simplicity), following the same criterion as in ref.~\cite{babusiaux2018} based on the tangential velocities of individual stars. Figure~\ref{fig:new_figure} shows the recovered SFHs for the thin (light blue) and thick (dark blue) discs, respectively. As expected, the thick disc presents a larger contribution of old stars than the thin disc and, although the youngest bursts are also visible in the thick disc they are clearly more prominent in the thin disc. However, the burst at $\sim$ 5.7 Gyr is quite prominent in both kinematic components (16$\%$ of the thick stars, 24$\%$ of the thin disc stars), suggesting that the event giving rise to such burst should have been more powerful and have a more global effect than those producing the more recent bursts (mainly observed in the thin disc). Pinpointing the causes of the found star formation episodes will boost our knowledge on our own Galaxy history, providing crucial information on the processes driving star formation in galactic scales.

Galaxy major mergers are thought to be one of the main factors triggering star formation in galaxies\cite{kennicutt1987, tissera2002, ellison2013}, and are predicted to have clear effects on their chemical evolution\cite{tinsley1979}. In particular, the current cosmological paradigm predicts that massive mergers play a crucial role in the formation of galaxies\cite{freeman2002ARAA}, including the Milky Way\cite{meza2005, gallart2019}. However, we do not have any indications of such massive mergers happening in the late evolution of our Galaxy\cite{freeman2002ARAA, naab2017}.

On the other hand, Milky Way halo substructures or stellar streams such as those linked to Sgr\cite{ibata1994, belokurov2006} do suggest that our Galaxy has experienced minor satellite encounters in the last few billion years. We compared the times at which we find star formation enhancements with those of the proposed first infall and following pericentre passages for Sgr from numerical simulations. The coincidence is striking (see Figure~\ref{fig:plot2}, panel c). Mimicking the observed angular positions, distances, and radial velocities of tidal streams from Sgr, simulations of the Sgr tidal disruption suggest close pericentric passages ($\sim$ 10 kpc) around 6.5, 4.5, 2.75, 1, and 0.1 Gyr ago\cite{law2010} (secondary pericentres at 5.5, 3.7, and 1.9 Gyr ago). Other attempts at reconstructing the orbital history of Sgr, some including the possible effect of the Magellanic Clouds, suggest that Sgr could have been affecting the Milky Way disc since 6-7 Gyr ago, dating the first Sgr infall around 5 Gyr ago, with subsequent passages also compatible with these observed SFH enhancements\cite{purcell2011, laporte2018}. Further constraints on the orbit of Sgr come from the observed peculiarities in the phase space and stellar streaming motions in the Milky Way disc\cite{eggen1969, siebert2011, williams2013}, which have been explained as the result of perturbations excited by pericentre passages of Sgr\cite{gomez2013}. Such observations can be explained if Sgr has a total mass of M$_{\rm total}$~$\sim$~2.5$\times$10$^{10}$~M$_{\odot}$ and passed close to the Milky Way disc 2.2 and 1.1 Gyr ago. These two recent pericentric passages also agree closely with two of the bursts of star formation derived with our techniques. New studies based on Gaia DR2 data have unveiled even more detailed kinematic features that suggest that the Galaxy is currently experiencing an ongoing phase-mixing process\cite{katz2018, antoja2018, laporte2019}. Comparison of the observed distribution of stars in the vertical position and velocity (Z-V$_{\rm Z}$) plane with N-body models of galaxy dynamics suggest that the Milky Way disc was perturbed just 300 to 900 million years ago, in reasonable agreement with the most recent Sgr pericentre passage\cite{law2010, purcell2011, delavega2015, laporte2018}. Such findings reinforce the idea of Sgr being the main perturbing agent during the last 6 Gyr of evolution of the Milky Way\cite{laporte2019}. In fact, a recent work combining APOGEE data with Gaia DR2 claims that satellite perturbations should be an important heating mechanism of the outer disc to explain the high vertical velocity dispersion of $\sim$~6~Gyr-old stars as well as the relationship between [Fe/H] and radial age-velocity dispersion\cite{mackereth2019}. According to our derived SFH, Sgr is also an important actor in the build up of the Milky Way disc stellar content.

So far, we have discussed only the effect of Sgr on the Milky Way. However, such interactions should have also had an effect on Sgr, which is significantly less massive than our Galaxy\cite{purcell2011}. Analyses on the stellar content of the Sgr stream via CMD fitting infer a typical dwarf-like star formation and chemical enrichment history until 5-7 Gyr ago\cite{deboer2015}. At that time, star formation in the Sgr stream halted, giving way to a residual activity. This drop in the star formation, together with the presence of the first enhancement on the SFH reported in this work, strongly suggest that the first infall of Sgr into the Milky Way potential took place around 5.5 Gyr ago. Such event forced the stripping of gas in Sgr, quenching star formation, and triggered the formation of new stars in the Milky Way due to induced shocks and perturbations, and possibly, gas accretion. However, the analysis of just the stream, associated with the outer parts of the original dwarf galaxy and most likely stripped during this first infall, disables the reconstruction of more recent events. Fortunately, high-quality Hubble Space Telescope data of the core of the Sgr main body is also available\cite{siegel2007}. Its CMD is characterised by the presence of several main-sequence turnoffs (the most robust age diagnostics), indicating distinct star forming epochs. Apart from a metal-poor population older than 8 Gyr (globular cluster like), careful examination of the stellar content of the Sgr core is indicative of three distinct stellar populations: i) a prominent and extended, intermediate-age population (4-8 Gyr old), with the bulk of stars being around 5-6 Gyr old; ii) a younger, slightly more metal-rich population of age 2.3 Gyr; and iii) a metal-rich (above solar), young population of $\sim$ 0.9 Gyr. The tight correlation between the stellar content of Sgr and the Milky Way disc SFH further hints at the hypothesis of the newly-reported bursts of star formation in the Milky Way being linked to interactions with Sgr. However, at this point, we should note that Sgr is not the only Milky Way satellite that might be currently affecting our Galaxy. Several works suggest that the Magellanic Clouds are in first infall toward the Milky Way\cite{kallivayalil2006, besla2007}. As a consequence, we cannot discard a possible effect of the Magellanic system on the Milky Way's SFH in the last $\sim$ 1-2 Gyr, and particularly around the time of the last enhancement $\sim$ 0.1 Gyr ago, when the first LMC pericentric approach may have occurred\cite{laporte2018LMC}.

It is worth noting that there might be a number of factors that could delay star formation linked to predicted pericentre passages. Among the possible reasons we can name the likely time lag between close passages and star formation enhancements or limitations on the dynamical models to accurately predict orbital histories (e.g. dynamical friction). Still, the episodic nature of the SFH recovered for the Milky Way disc clearly implies that drastic events should have occurred around 6, 2 and 1 Gyr ago. As detailed above, a wealth of evidence pile up indicating that Sgr (first infall and subsequent pericentric passages) might be behind the reported star formation enhancements. These results perfectly supplement a recent characterisation of the stellar content of the Anticentre Stream, where claims are made that this stream is the outcome of the interaction between the Milky Way and Sgr in its first infall\cite{laportbelokurov2019}. Indeed, of all pericentric passages of Sgr, first infall is the one that seems to affect more globally the SFH of the Milky Way disc. Figure~\ref{fig:new_figure} clearly shows that, while the first Sgr passage considerably triggered star formation in both disc kinematic components, subsequent passages were only able to produce significant star formation in the thin disc.

The possibility that a satellite dwarf galaxy remarkably less massive than its host is able to induce repeated massive events of star formation involving a significant part of its disc is striking. Star formation induced by tidal perturbations has been reported before in galaxy pairs with lower mass ratio. 
An example in our vicinity is that of the M31-M33 pair (mass ratio $\sim$~7:1). N-body simulations indicate that M31 and M33 had their latest pericentric passage $\sim$~2.6~Gyr ago\cite{mcconnachie2009}, and the analysis of deep Hubble Space Telescope CMDs\cite{bernard2012} reveals a clear episode of star formation peaking $\sim$~2~Gyr ago, in both M31 and M33 outer regions, in close agreement with the timing of the suggested pericentre. In more distant galaxies, only ongoing star formation bursts associated with current interactions can be reasonably identified due to the difficulty to obtain detailed time resolved SFHs for these systems\cite{kennicutt1987}.  

Theoretical works have historically asserted that relatively massive mergers should indeed enhance star formation in the embroiled systems\cite{mihos1994, hernquist1995, naab2017}. In particular, star forming bursts have been found within a cosmological framework in interacting systems with mass ratios up to $\sim$~10:1\cite{tissera2002}. Some other works claim that the star formation efficiency should decrease at high mass ratios, and indicate that little enhancement of star formation can be expected at the Sgr-Milky Way mass ratio\cite{cox2008}. Other models of the effect of mergers on the SFH of interacting galaxies\cite{mihos1994, hernquist1995} indicate that the induced star formation should be concentrated to the inner parts, as a consequence of gas infall in response to tidal forces. This is expected even in the case of minor mergers (mass ratio $\sim$~10:1 or lower).
Some other simulations of major mergers (up to 2.5:1) even find that such star formation enhancements in central regions are accompanied by a moderate suppression of star formation at larger galactocentric distances\cite{moreno2015}, in clear contradiction to the global star formation episodes that we find.

However, modern attempts, including more complex and detailed modelling of star formation, are starting to cast doubts on these ideas. 
A proper understanding of the underlying physical processes behind star formation in galaxies and its inclusion in models is one of the major challenges for galaxy formation and evolution theory\cite{naab2017}. The increasingly detailed modelling of the interstellar medium, including refinements such as multi-phase structure, has provided results that have begun to change our understanding of star formation episodes in merging systems. Rather than inducing large-scale gas inflows towards the central parts, certain mergers can result in gas fragmentation into massive and dense clouds\cite{teyssier2010}. Other aspects such as modification of the modelling of star formation itself (based on density or shock-induced) and the effect of a hot gaseous halo (usually not included), also influence the way star formation is enhanced in simulations of interacting galaxies\cite{chien2010, moster2011}. In all cases, implementation of a more realistic and higher resolution modelling affect the starburst efficiency and, more importantly for this discussion, result in star formation enhancements in regions at larger galactocentric radius as opposed to the centrally concentrated star formation suggested by previous studies. Unfortunately, to our knowledge, there are no theoretical works where a single satellite could induce several global star formation enhancements as Sgr seems to be causing in the Milky Way disc. Therefore, although we can easily link the reported enhancements with possible pericentric passages of Sgr, we cannot pinpoint what exact physical mechanisms are triggering such events.

\vspace{0.5cm}

In this work we obtained the most detailed SFH of the 2-kpc local volume to date. Strikingly, we find, for the first time, that such volume is characterised by an episodic SFH, with clear enhancements of star formation $\sim$ 5.7, 1.9, and 1 Gyr ago. All evidence seems to suggest that recurrent interactions between the Milky Way and Sgr dwarf galaxy are behind such enhancements. These findings imply that low mass satellites, not only affect the Milky Way disc dynamics, but also are able to trigger significant events of star formation throughout its disc. The precise dating of such star forming episodes provided in this work sets useful boundary conditions to properly model the orbit of Sgr and its interaction with the Milky Way. In addition, this work can be understood as a major milestone, providing important constraints on the modelling of the interstellar medium and star formation within hydrodynamical simulations, manifesting the need of understanding physical processes at subresolution scales and of further analysis to unveil the physical mechanisms behind global and repeated star formation events induced by satellite interaction.


\vspace{0.25cm}

\noindent
{\bf Acknowledgements} The authors are grateful to the anonymous referees for their invaluable job at improving the original manuscript. We would like to thank Kenneth C. Freeman, Coryn Bailer-Jones, Giuseppina Battaglia, Mike Beasley, Chris Brook, Claudio Dalla Vecchia, Jes\'us Falc\'on-Barroso, Ryan Leaman, and Isabel P\'erez for very useful discussions. TRL and CG acknowledge financial support through the grants (AEI/FEDER, UE) AYA2017-89076-P, AYA2016-77237-C3-1-P (RAVET project) and AYA2015-63810-P, as well as by the Ministerio de Ciencia, Innovaci\'on y Universidades (MCIU), through the State Budget and by the Consejer\'\i a de Econom\'\i a, Industria, Comercio y Conocimiento of the Canary Islands Autonomous Community, through the Regional Budget (including IAC project, TRACES). TRL is supported by a MCIU Juan de la Cierva - Formaci\'on grant (FJCI-2016-30342). SC acknowledges support from Premiale INAF ``MITIC'' and grant AYA2013-42781P from the Ministry of Economy and Competitiveness of Spain, he has also been supported by INFN (Iniziativa specifica TAsP). We used data from the European Space Agency mission Gaia (\url{http://www.cosmos.esa.int/gaia}), processed by the Gaia Data Processing and Analysis Consortium (DPAC; see \url{http://www.cosmos.esa.int/web/gaia/dpac/consortium}). Funding for DPAC has been provided by national institutions, in particular the institutions participating in the Gaia Multilateral Agreement.

\vspace{0.25cm}

\noindent
{\bf Author contributions} The writing of the manuscript was mainly carried out by TRL together with CG. TRL and CG defined the final samples under analysis and extracted the star formation histories presented in this work. The software to analyse Gaia DR2 data was written by TRL and EJB. SC contributed with the tools for generating the synthetic CMDs as well as with evolutionary model predictions in the Gaia photometric system. TRL, CG, EJB, SC contributed to the interpretation and analysis of the results.

\vspace{0.25cm}

\noindent
{\bf Competing interests} The authors declare no competing financial interests

\vspace{0.25cm}

\noindent
{\bf Correspondence and requests for materials} should be addressed to TRL ({\it tomasruizlara@gmail.com}).

\section*{Methods}

The European Space Agency (ESA) mission Gaia is providing an unprecedented view of our immediate cosmic neighbourhood through the mapping of the stellar content of vast volumes around the Milky Way. In particular, the precise parallaxes to individual stars delivered in Gaia DR2\cite{brown2018} are among the most relevant breakthroughs. They allow obtaining deep CMDs in the absolute magnitude plane for large volumes within the Milky Way for the first time\cite{babusiaux2018}. The analysis of such CMDs enable the computation of detailed and quantitative SFHs\cite{gallart2005} extended from the very earliest times in the Milky Way evolution to the present. We should emphasise here that the quality and volume of the Gaia data in combination with advanced analysis techniques and a robust theoretical framework are crucial and essential ingredients to unveil the SFH of the Milky Way with an unprecedented detail and resolution.

\section{Gaia data and sample selection}

We make use of Gaia DR2 data\cite{brown2018} to infer the SFH characteristic of the disc of the Milky Way using the well-tested technique of CMD fitting. The original sample is composed by all Gaia DR2 stars within a sphere of $\sim$ 2 kpc of radius (distance obtained by simply inverting the parallax, i.e. parallax $\geq$ 0.5, not considering parallax offset yet, see below) and brighter than M$_G$ = 7 (absolute magnitude). From the nearly 82 million stars fulfilling these initial criteria we then restrict to those displaying a relative uncertainty in the parallax below 20$\rm \%$ (\verb|parallax_over_error| $>$ 5\cite{luri2018}, keeping around 90$\%$ of the stars). The analysed sample of stars is drawn from this initial sample after some extra technical aspects are taken into account:
\begin{itemize}

\item The parallaxes provided by the Gaia Collaboration are known to be underestimated\cite{lindegren2018}. Several works in the literature studying the Gaia DR2 parallax zero-point suggested different values ranging from 0.029 to 0.082 mas (refs.\cite{stassun2018, riess2018, zinn2018, schoenrich2019, khan2019, graczyk2019, hall2019, leung2019}). We corrected the parallaxes of each and every star in our sample using a single value of 0.054 mas, not considering possible spatial variations, following ref.\cite{schoenrich2019}. We have repeated the whole analysis using parallax offset values near the extremes of the proposed range, namely 0.029 and 0.082 mas. This has allowed us to verify that the particular parallax offset adopted within reasonable values has little impact in the results presented in this work, although the percentage of old stars tends to be higher when using large values (i.e. 0.082 mas).
\item After considering this parallax offset, we have computed absolute magnitudes using the inverse of the parallax as a distance estimator to individual stars\cite{gallart2019}. Different works in the literature provide Bayesian geometric distance estimates\cite{bailerjones2018, schoenrich2019, anders2019} (among other quantities) refining the Gaia DR2 data products that are pertinent for stellar population analysis. Although highly advisable for studies involving stars at large distances from the Sun, given the large parallaxes of our sources and their small parallax errors, the simple approximation followed in this work to compute distances is perfectly valid (see figure 6 in ref.\cite{bailerjones2018}, figure 13 in ref.\cite{schoenrich2019}, or figure 18 in ref.\cite{anders2019}, to name but a few).
\item We correct the Gaia magnitudes for interstellar extinction interpolating the colour excess from a 3-D map valid in the analysed volume\cite{lallement2018}. Different extinction coefficients are used\cite{casagrande2018, babusiaux2018} with little or no effect on the final results (see figure~\ref{fig:plot2}). SFHs represented by solid lines are computed using the Casagrande coefficients\cite{casagrande2018}, while those depicted using dashed lines make use of the Babusiaux coefficients\cite{babusiaux2018}. In the case of the Casagrande coefficients, temperatures for individual stars were estimated using a correlation between G$_{\rm BP}$-G$_{\rm RP}$ colour and effective temperature determined from a large sample of Gaia DR2 stars with measured temperatures (dependence with metallicity was not considered). Highly reddened stars (A$_{\rm G} > $ 0.5) are not included in the final sample (40$\%$ of the stars are removed by applying this criterion). This cut introduces a spatial bias in the sample, since stars close to the plane and at larger distances from the Sun are affected by a larger amount of reddening. The inset of Figure~\ref{fig:new_figure} shows the spatial location of the stars in the final sample. From the different SFHs for the kinematic thin and thick disks shown in that Figure, it can be deduced that the spatial bias introduced by the reddening cut likely results in somewhat underestimated values of the young star formation rate, possibly also underestimating the heights of the youngest bursts.
\item Given our knowledge on the stellar content of the Milky Way, we expect that its oldest Main Sequence Turn-Off is located at around M$_G$ = 4. To ensure that we reach at least 1 magnitude deeper than this critical position, after the parallax offset is taken into account and the magnitudes are corrected by interstellar extinction, we keep for further analysis only those stars brighter than M$_G$ = 5. In this step almost half of the stars are removed.
\item Finally, following refs.\cite{babusiaux2018, lindegren2018}, stars not fulfilling the following criterion were discarded (less than 1$\%$ of the stars): \\
\verb|astrometric_chi2_al|/(\verb|astrometric_n_good_obs_al| - 5) $<$ \\
 $<$ 1.44$\rm \times$max(1, exp(-0.4$\rm \times$ (\verb|phot_g_mean_mag|-19.5))) \\
Finally, stars with no G$_{\rm BP}$-G$_{\rm RP}$ colour determination available are by definition not included in our sample (less than 0.2 $\%$).
\end{itemize}

At this point, we should highlight the fact that our sample is essentially complete, as all the analysed stars are brighter than G = 17.0 (apparent magnitude, taking into account reddening\footnote{An absolute magnitude of M$_G$ = 5, at a maximum of 2 kpc, corresponds to an apparent magnitude of $\sim$ 16.5}). In the brightest end (G $<$ 12), Gaia DR2 is estimated to be 97$\%$ complete\cite{mor2019}, so our sample is basically complete also at the bright end. This is important as any incompleteness or bias in the coverage of the observed CMD could affect the results and thus, should be properly replicated in the synthetic CMD, which is the essential ingredient in the CMD fitting technique that we use in this work (see next section).



\section{Star formation histories from Gaia colour-magnitude diagrams}

The analysis of deep CMDs has long been used and demonstrated as an excellent way to study SFHs in dwarf galaxies within the Local Group\cite{gallart2005, cignonitosi2010, apariciohidalgo2009, tolstoy2009, cignoni2010}. In this work, we derive SFHs of the Milky Way disc by comparing Gaia DR2 observed CMDs to synthetic ones following the methodology carefully outlined in ref.\cite{monelli2010}. 

We used a synthetic CMD containing 200 million stars with ages and metallicities ranging from 0.03 to 14 Gyr and 0.0001 to 0.05, respectively. This synthetic CMD was computed using the BaSTI stellar evolution library\cite{pietrinferni2004} in its solar-scaled version, with a Reimers mass loss parameter ($\eta$) of 0.2, assuming a Kroupa initial mass fraction\cite{kroupa2001}, a binary fraction of 70\% ($\beta$), and a minimum mass ratio for binaries of 0.1 (q). Regarding the treatment of binaries, we should clarify here that the stars in a binary system are drawn from the same (single star) stellar evolutionary tracks, and their luminosities are combined. This should mimick the fact that Gaia DR2 contains some non-resolved binary systems.

The predictions from the stellar evolutionary models were translated to the Gaia photometric system by using bolometric correction tables following ref.~\cite{evans2018} for Gaia DR2 (revised passbands corresponding to the data plotted in their figure 21 --ascii data downloaded from the official Gaia home page-- and the corresponding zero points in their table 1 column 4). Before comparing observed with synthetic CMDs, we need to simulate the effect of observational errors in the synthetic diagrams following the procedure explained in ref.\cite{gallart2019}. Given the small photometric errors of Gaia \cite{evans2018}, the major sources of uncertainty in the magnitude and colour of the stars stem from the error in the distance estimation used to transform from apparent to absolute magnitudes using the parallax, and from the errors in the reddening correction (which dominate the error in the G$_{\rm BP}$-G$_{RP}$ colour). We mimic  observational errors in M${_{\rm G}}$ on a star-by-star basis using the equation $\rm \Delta$M$_{\rm G}$ = 2.17/\verb|parallax_over_error| (see ref.\cite{gallart2019} for more information). To this end, we recreate the observed \verb|parallax_over_error| distribution in the synthetic CMD (including parallax offset). Then, each theoretical M${_{\rm G}}$ magnitude is modified by adding a correction following a Gaussian distribution centred at 0 with a standard deviation of $\rm \Delta$M$_{\rm G}$. Small errors in the Gaia photometry itself \cite{evans2018} as well as in the reddening correction, are introduced in a similar fashion by adding a dispersion of 0.01 in the G$_{\rm BP}$-G$_{RP}$ colour (see ref.~\cite{evans2018} and especially ref.~\cite{gallart2019} for further details).

The comparison itself was done using the code ``THESTORM'' (``Tracing tHe Evolution of the STar fOrmation Rate and Metallicity'')\cite{bernard2015letter, bernard2018MNRAS}. In this step, the combination of simple stellar populations (SSP, with a small range of age and metallicity) that best fits the observed CMD is determined. These SSPs are drawn from the synthetic CMD after simulation of observational errors.  The comparison is carried out by counting the number of stars in colour-magnitude boxes of different sizes depending on their location within the CMD, where a number of {\it bundles} have been defined (see Figure~\ref{fig:plot1}). Bundles are regions in the CMD sampled with an homogeneous box size, chosen according to the reliability of the stellar evolution models in that region, or the density of stars in them. We have shown that the exact bundle configuration has little effect in the final SFH\cite{ruizlara2018}. THESTORM simultaneously characterizes the stellar content, in age and metallicity, of the sample under analysis not considering any a-priori constraint (i.e. no initial shape of the star formation rate as a function of time, or age-metallicity relation is adopted). Uncertainties on the SFH were computed as described in ref.\cite{hidalgo2011}. They are assumed to be a combination in quadrature of the uncertainties due to the effect of binning in the colour–magnitude and age–metallicity planes (from the dispersion of different solutions obtained after shifting the bin limits), and those due to the effect of statistical sampling in the observed CMD (the dispersion of several solutions obtained after resampling the observed CMD following Poissonian statistics). The applied methodology in its current version does not take into account the effect on the error computation of secondary factors such as a possible degeneracy in the solution due to secondary solutions (close to the best fit), or the choice of the stellar population parameters adopted to calculate the synthetic CMD (initial mass fraction, binary simulation, etc.), which do not exactly match those of the observed population. For a more detailed explanation on the fitting procedure and error calculation, we refer the reader to refs.\cite{monelli2010, hidalgo2011, bernard2018MNRAS}.



For a proper computation of the SFH coming from a given sample of observed stars, the synthetic CMD to which we compare should greatly outnumber the observed stars. For this reason, and to enable the computation of the SFH of such a vast volume ($\sim$ 24 million stars in the $\sim$ 2 kpc radius sphere, after quality cuts have been applied), we randomly select several subsamples from the original one, containing up to $\sim$~250000 stars. Figure~\ref{fig:plot1} shows one of these fits (panel b) compared with the observed CMD (panel a). Panel d shows  the residual difference between the observed and the best model CMD in poissonian sigmas. Note that in the great majority of the CMD the residuals are lower that 2 sigmas. The largest difference between the model and the observed CMD are observed in the red clump and in the fainter part of the main sequence. However, these differences are usually within $2-3\sigma$ and have understood origins. i) The presence of higher than average residuals in the region of the CMD corresponding to the core He-burning stage, is mainly due to the fact that the adopted version of the BaSTI library relies on the use of conductive opacity prescriptions that have been improved in the new BaSTI library\cite{hidalgo2018, cassisi2007} and that have the effect of slightly decreasing the brightness of the horizontal branch stellar models, which would contribute to improve the match between the synthetic and the observed CMD; additionally, the exact morphology - in particular the colour distribution - of the red clump strongly depends on the adopted mass loss efficiency during the previous RGB stage, for which we have adopted a (fixed) average efficiency. ii) In the case of the main sequence below the old main sequence turnoff, the positive residuals are related to the stars for which the reddening correction was underestimated, and that, therefore, could not be reproduced in the model. These stars are shifted to a redder position, and thus are absent from their true, blue position in the main sequence (thus the negative residual there). Figure~\ref{fig:plot2} shows the recovered SFH for all the trials we have performed. Different lines represent the SFHs from the analysis of different random selections and different extinction coefficients to ensure the robustness of the result and the negligible effect of reducing the sample under analysis. The final solution (orange line in figure~\ref{fig:plot2}) is computed after averaging all the solutions. Error bars encompassed the 16$^{th}$ and 84$^{th}$ percentile of all solutions.

\begin{figure}[h]
\begin{center}
\includegraphics[width=0.95\textwidth]{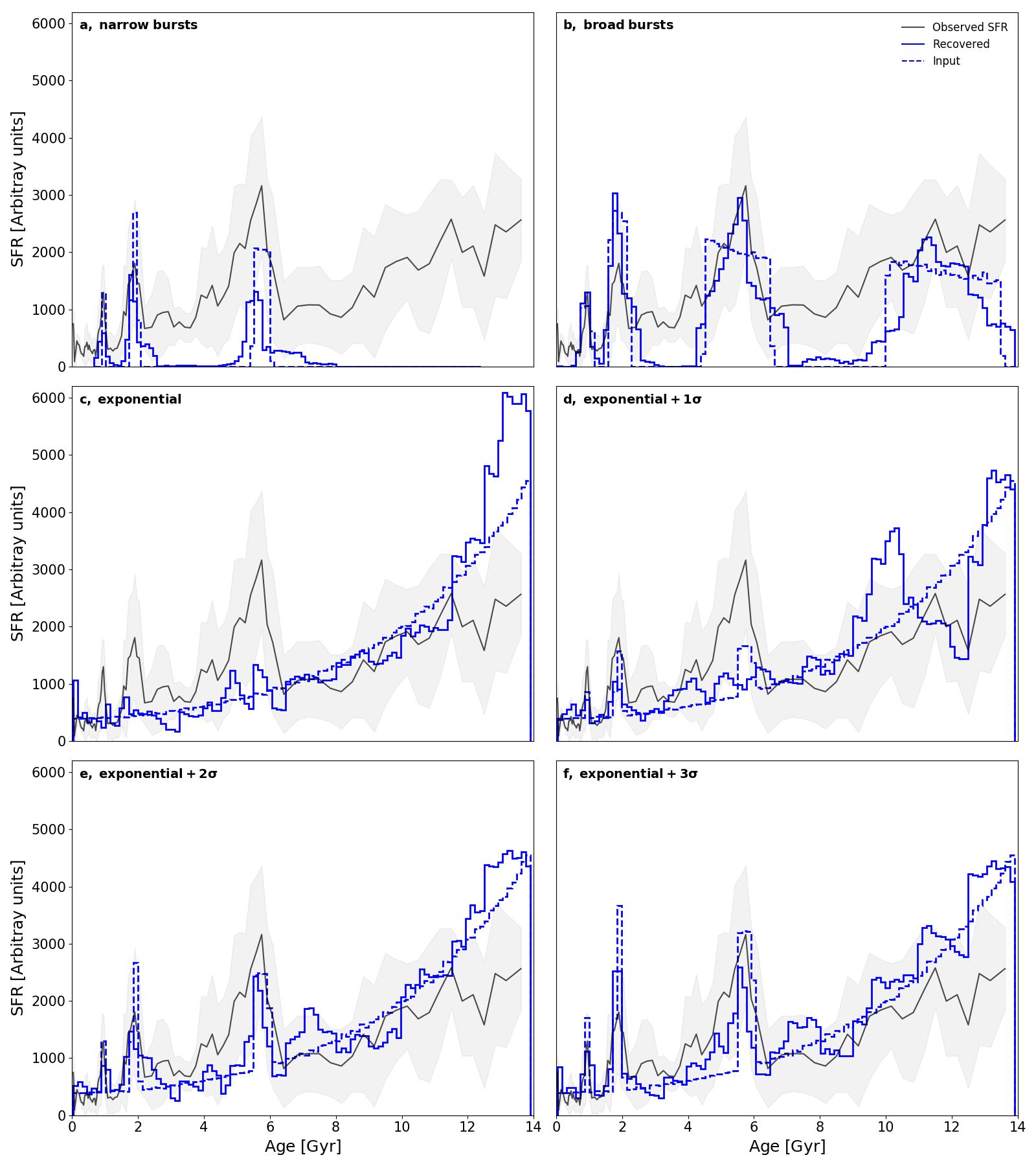} 
\end{center}
\caption{{\bf Testing the robustness of the SFH recovery (ii)}. In this figure we show the recovered SFH for the 2 kpc volume analysed from Gaia (grey), the input SFH for the mock CMD (dashed blue line) and the recovered SFH after mimicking errors and applying our methodology (solid blue line). Top row: only narrow (left) and wide (right) bursts. Middle and bottom row: Exponential decline plus bursts of different strengths (0 to 3$\sigma$).}
\label{fig:plot6}
\end{figure}

The sensitivity of this method to accurately recover the age and duration of well-defined and isolated bursts of star formation has been extensively tested\cite{hidalgo2011}, reinforcing our results. Nevertheless, we have performed extra tests for the particular case of Gaia data focusing on the results presented here. We created a series of mock stellar populations, i.e. a set of $\sim$ 200000 synthetic stars from the BaSTI models following a particular SFH (exponential decline, burst of star formation and combinations of exponential decline and bursts of different widths and strengths, etc.). We then simulate the observational errors to such synthetic mock CMDs as explained above to finally obtain a recovered SFH using ``THESTORM'' (same methodology as the one described above). This experiment is shown in Figure~\ref{fig:plot6}. We can conclude that the bursts that we recover for the 2-kpc volume around the solar neighbourhood (see Figure~\ref{fig:plot2}) are significant above 3-4$\sigma$ and that the oldest burst is more extended than the younger ones.

\section{Data availability}
 
The data that support the plots within this paper and the findings of this study are available from the corresponding author upon reasonable request.

\end{document}